# The KM3NeT multi-PMT optical module

---

S. Aiello, [a] A. Albert, [bb,b] M. Alshamsi, [c] S. Alves Garre, [d] Z. Aly, [e] A. Ambrosone, [f,g] F. Ameli, [h] M. Andre, [i] G. Androulakis [j,1] M. Anghinolfi, [k] M. Anguita, [l] M. Ardid, [m] S. Ardid, [m] J. Aublin, [c] T. Avgitas, [c] C. Bagatelas, [j] L. Bailly-Salins, [n] B. Baret, [c] S. Basegmez du Pree, [o] M. Bendahman, [c,p] F. Benfenati, [q,r] E. Berbee, [o] A. M. van den Berg, [s] V. Bertin, [e] V. van Beveren, [o] S. Biagi, [t] R. de Boer, [o] M. Boettcher, [u] M. Bou Cabo, [v] J. Boumaaza, [p] M. Bouta, [w] M. Bouwhuis, [o] C. Bozza, [x] H. Brânzaş, [y] R. Bruijn, [o,z] J. Brunner, [e] R. Bruno, [a] E. Buis, [o,aa] R. Buompane, [f,ab] J. Busto, [e] G. Cacopardo, [t] B. Caiffi, [k] D. Calvo, [d] S. Campion, [ac,h] A. Capone, [h] V. Carretero, [d] P. Castaldi, [q,ad] S. Celli, [ac,h] M. Chabab, [ae] C. Champion, [c] N. Chau, [c] A. Chen, [af] S. Cherubini, [t,ag] V. Chiarella, [ah] T. Chiarusi, [q] M. Circella, [ai] R. Cocimano, [t] J.A.B. Coelho, [c] A. Coleiro, [c] M. Colomer Molla, [c,d] R. Coniglione, [t] P. Coyle, [e] A. Creusot, [c] A. Cruz, [aj] G. Cuttone, [t] C. D'Amato, [t] A. D'Amico, [o] R. Dallier, [ak] A. De Benedittis, [f] B. De Martino, [e] G. De Wasseige, [bc] I. Di Palma, [ac,h] A.F. Díaz, [l] D. Diego-Tortosa, [m] C. Distefano, [t] A. Domi, [o,z] C. Donzaud, [c] D. Dornic, [e] M. Dörr, [al] E. Drakopoulou, [j] D. Drouhin, [bb,b] T. Eberl, [am] A. Eddyamoui, [p] T. van Eeden, [o] D. van Eijk, [o] I. El Bojaddaini, [w] S. El Hedri, [c] A. Enzenhöfer, [e] V. Espinosa, [m] P. Fermani, [ac,h] G. Ferrara, [t,ag] M. D. Filipović, [an] F. Filippini, [q,r] J. Fransen, [o] L.A. Fusco, [x] D. Gajanana, [o] T. Gal, [am] J. García Méndez, [m] A. García Soto, [d] F. Garufi, [f,g] Y. Gatelet, [c] C. Gatius Oliver, [o] N. Geißelbrecht, [am] L. Gialanella, [f,ab] E. Giorgio, [t] S.R. Gozzini, [d] R. Gracia, [am] K. Graf, [am] G. Grella, [ao] A. Grmek, [t] I. Gromov, [o] D. Guderian, [bd] C. Guidi, [k,ap] B. Guillon, [aq] M. Gutiérrez, [ar] J. Häfner, [am] L. Haegel, [c] S. Hallmann, [am] H. Hamdaoui, [p] H. van Haren, [as] A. Heijboer, [o] A. Hekalo, [al] L. Hennig, [am] J.J. Hernández-Rey, [d] J. Hofestädt, [am] F. Huang, [e] W. Idrissi Ibnsalih, [f,ab] A. Ilioni, [c] G. Illuminati, [q,c,r] C.W. James, [aj] D. Janezashvili, [at] P. Jansweijer, [o] M. de Jong, [o,au] P. de Jong, [o,z] B.J. Jung, [o] P. Kalaczyński, [av] O. Kalekin, [am] U.F. Katz, [am] F. Kayzel, [o] N.R. Khan Chowdhury, [d] G. Kistauri, [at] F. van der Knaap, [aa] P. Kooijman, [z,be] J. Koopstra, [o] A. Kouchner, [c,aw] V. Kulikovskiy, [k] M. Labalme, [aq] R. Lahmann, [am] M. Lamoureux, [c,2] G. Larosa, [t] C. Lastoria, [e] A. Lazo, [d] R. Le Breton, [c] S. Le Stum, [e] G. Lehaut, [aq] O. Leonardi, [t] F. Leone, [t,ag] E. Leonora, [a] N. Lessing, [am] G. Levi, [q,r] M. Lincetto, [e] M. Lindsey Clark, [c] C. LLorens Alvarez, [m] F. Longhitano, [a] D. Lopez-Coto, [ar] L. Maderer, [c] J. Majumdar, [o] J. Mańczak, [d] A. Margiotta, [q,r] A. Marinelli, [f] C. Markou, [j] L. Martin, [ak] J.A. Martínez-Mora, [m] A. Martini, [ah] F. Marzaioli, [f,ab] S. Mastroianni, [f] K.W. Melis, [o] G. Miele, [f,g] P. Migliozzi, [f] E. Migneco, [t] P. Mijakowski, [av] L.S. Miranda, [ax] C.M. Mollo, [f] M. Mongelli, [ai] A. Moussa, [w] R. Muller, [o] M. Musumeci, [t] L. Nauta, [o] S. Navas, [ar] C.A. Nicolau, [h] B. Nkosi, [af] B. Ó Fearraigh, [o,z] M. O'Sullivan, [aj] M. Organokov, [b] A. Orlando, [t] J. Palacios González, [d] G. Papalashvili, [at] R. Papaleo, [t] G. Passaro [Gianluca], [t] G. Passaro [Giuseppe], [t] C. Pastore, [ai] A. M. Păun, [y] G.E. Păvălaş, [y] G. Pellegrini, [q] C. Pellegrino, [r,bf] M. Perrin-Terrin, [e] V. Pestel, [o] P. Piattelli, [t] C. Pieterse, [d] O. Pisanti, [f,g] C. Poirè, [m] V. Popa, [y] T. Pradier, [b] I. Probst, [am] S. Pulvirenti, [t] G. Quéméner, [aq] N. Randazzo, [a] S. Razzaque, [ax] D. Real, [d] S. Reck, [am] G. Riccobene, [t] R. Rocco, [f] A. Romanov, [k,ap] A. Rovelli, [t] F. Salesa Greus, [d] D.F.E. Samtleben, [o,au] A. Sánchez Losa, [ai,d] M. Sanguineti, [k,ap] D. Santonocito, [t] P. Sapienza, [t] R. Schmeitz, [o] J. Schmelling, [o] J. Schnabel, [am] M.F. Schneider, [am] J. Schumann, [am] H. M. Schutte, [u] J. Seneca, [o] I. Sgura, [ai] R. Shanidze, [at] A. Sharma, [ay] A. Sinopoulou, [j] M.V. Smirnov, [am] B. Spisso, [ao,f] M. Spurio, [q,r] D. Stavropoulos, [j] S.M. Stellacci, [ao,f] M. Taiuti, [k,ap] F. Tatone, [ai] Y. Tayalati, [p] H. Tedjditi, [k] H. Thiersen, [u] P. Timmer, [o] S. Tingay, [aj] S. Tsagkli, [j] V. Tsourapis, [j] E. Tzamariudaki, [j] D. Tzanetatos, [j] C. Valieri, [q] V. Van Elewyck, [c,aw] G. Vasileiadis, [az] F. Versari, [q,r] S. Viola, [t] D. Vivolo, [f,ab] J. Wilms, [ba] R. Wojaczyński, [av] E. de Wolf, [o,z] T. Yousfi, [w] S. Zavatarelli, [k] A. Zegarelli, [ac,h] D. Zito, [t] J.D. Zornoza, [d] J. Zúñiga, [d] and N. Zywucka, [u]

[a] INFN, Sezione di Catania, Via Santa Sofia 64, Catania, 95123 Italy
[b] Université de Strasbourg, CNRS, IPHC UMR 7178, F-67000 Strasbourg, France
[c] Université de Paris, CNRS, Astroparticule et Cosmologie, F-75013 Paris, France
[d] IFIC - Instituto de Física Corpuscular (CSIC - Universitat de València), c/Catedrático José Beltrán, 2, 46980 Paterna, Valencia, Spain
[e] Aix Marseille Univ, CNRS/IN2P3, CPPM, Marseille, France
[f] INFN, Sezione di Napoli, Complesso Universitario di Monte S. Angelo, Via Cintia ed. G, Napoli, 80126 Italy
[g] Università di Napoli "Federico II", Dip. Scienze Fisiche "E. Pancini", Complesso Universitario di Monte S. Angelo, Via Cintia ed. G, Napoli, 80126 Italy
[h] INFN, Sezione di Roma, Piazzale Aldo Moro 2, Roma, 00185 Italy
[i] Universitat Politècnica de Catalunya, Laboratori d'Aplicacions Bioacústiques, Centre Tecnol{\`o
[j] NCSR Demokritos, Institute of Nuclear and Particle Physics, Ag. Paraskevi Attikis, Athens, 15310 Greece
[k] INFN, Sezione di Genova, Via Dodecaneso 33, Genova, 16146 Italy
[l] University of Granada, Dept. of Computer Architecture and Technology/CITIC, 18071 Granada, Spain
[m] Universitat Politècnica de València, Instituto de Investigación para la Gestión Integrada de las Zonas Costeras, C/ Paranimf, 1, Gandia, 46730 Spain
[n] LPC, Campus des Cézeaux 24, avenue des Landais BP 80026, Aubière Cedex, 63171 France
[o] Nikhef, National Institute for Subatomic Physics, PO Box 41882, Amsterdam, 1009 DB Netherlands
[p] University Mohammed V in Rabat, Faculty of Sciences, 4 av. Ibn Battouta, B.P. 1014, R.P. 10000 Rabat, Morocco
[q] INFN, Sezione di Bologna, v.le C. Berti-Pichat, 6/2, Bologna, 40127 Italy
[r] Università di Bologna, Dipartimento di Fisica e Astronomia, v.le C. Berti-Pichat, 6/2, Bologna, 40127 Italy
[s] KVI-CART University of Groningen, Groningen, the Netherlands
[t] INFN, Laboratori Nazionali del Sud, Via S. Sofia 62, Catania, 95123 Italy
[u] North-West University, Centre for Space Research, Private Bag X6001, Potchefstroom, 2520 South Africa
[v] Instituto Español de Oceanografía, Unidad Mixta IEO-UPV, C/ Paranimf, 1, Gandia, 46730 Spain
[w] University Mohammed I, Faculty of Sciences, BV Mohammed VI, B.P. 717, R.P. 60000 Oujda, Morocco
[x] Università di Salerno e INFN Gruppo Collegato di Salerno, Dipartimento di Matematica, Via Giovanni Paolo II 132, Fisciano, 84084 Italy
[y] ISS, Atomistilor 409, Măgurele, RO-077125 Romania
[z] University of Amsterdam, Institute of Physics/IHEF, PO Box 94216, Amsterdam, 1090 GE Netherlands
[aa] TNO, Technical Sciences, PO Box 155, Delft, 2600 AD Netherlands
[ab] Università degli Studi della Campania "Luigi Vanvitelli", Dipartimento di Matematica e Fisica, viale Lincoln 5, Caserta, 81100 Italy
[ac] Università La Sapienza, Dipartimento di Fisica, Piazzale Aldo Moro 2, Roma, 00185 Italy
[ad] Università di Bologna, Dipartimento di Ingegneria dell'Energia Elettrica e dell'Informazione "Guglielmo Marconi", Via dell'Università 50, Cesena, 47521 Italia
[ae] Cadi Ayyad University, Physics Department, Faculty of Science Semlalia, Av. My Abdellah, P.O.B. 2390, Marrakech, 40000 Morocco
[af] University of the Witwatersrand, School of Physics, Private Bag 3, Johannesburg, Wits 2050 South Africa
[ag] Università di Catania, Dipartimento di Fisica e Astronomia "Ettore Majorana", Via Santa Sofia 64, Catania, 95123 Italy
[ah] INFN, LNF, Via Enrico Fermi, 40, Frascati, 00044 Italy
[ai] INFN, Sezione di Bari, via Orabona, 4, Bari, 70125 Italy
[aj] International Centre for Radio Astronomy Research, Curtin University, Bentley, WA 6102, Australia
[ak] Subatech, IMT Atlantique, IN2P3-CNRS, Université de Nantes, 4 rue Alfred Kastler - La Chantrerie, Nantes, BP 20722 44307 France
[al] University Würzburg, Emil-Fischer-Straße 31, Würzburg, 97074 Germany
[am] Friedrich-Alexander-Universität Erlangen-Nürnberg (FAU), Erlangen Centre for Astroparticle Physics, Erwin-Rommel-Straße 1, 91058 Erlangen, Germany

[an] *Western Sydney University, School of Computing, Engineering and Mathematics, Locked Bag 1797, Penrith, NSW 2751 Australia*
[ao] *Università di Salerno e INFN Gruppo Collegato di Salerno, Dipartimento di Fisica, Via Giovanni Paolo II 132, Fisciano, 84084 Italy*
[ap] *Università di Genova, Via Dodecaneso 33, Genova, 16146 Italy*
[aq] *Normandie Univ, ENSICAEN, UNICAEN, CNRS/IN2P3, LPC Caen, LPCCAEN, 6 boulevard Maréchal Juin, Caen, 14050 France*
[ar] *University of Granada, Dpto. de Fìsica Teórica y del Cosmos & C.A.F.P.E., 18071 Granada, Spain*
[as] *NIOZ (Royal Netherlands Institute for Sea Research), PO Box 59, Den Burg, Texel, 1790 AB, the Netherlands*
[at] *Tbilisi State University, Department of Physics, 3, Chavchavadze Ave., Tbilisi, 0179 Georgia*
[au] *Leiden University, Leiden Institute of Physics, PO Box 9504, Leiden, 2300 RA Netherlands*
[av] *National Centre for Nuclear Research, 02-093 Warsaw, Poland*
[aw] *Institut Universitaire de France, 1 rue Descartes, Paris, 75005 France*
[ax] *University of Johannesburg, Department Physics, PO Box 524, Auckland Park, 2006 South Africa*
[ay] *Università di Pisa, Dipartimento di Fisica, Largo Bruno Pontecorvo 3, Pisa, 56127 Italy*
[az] *Laboratoire Univers et Particules de Montpellier, Place Eugène Bataillon - CC 72, Montpellier Cédex 05, 34095 France*
[ba] *Friedrich-Alexander-Universität Erlangen-Nürnberg (FAU), Remeis Sternwarte, Sternwartstraße 7, 96049 Bamberg, Germany*
[bb] *Université de Haute Alsace, rue des Frères Lumière, 68093 Mulhouse Cedex, France*
[bc] *UCLouvain, Centre for Cosmology, Particle Physics and Phenomenology, Chemin du Cyclotron, 2, Louvain-la-Neuve, 1349 Belgium*
[bd] *University of Münster, Institut für Kernphysik, Wilhelm-Klemm-Str. 9, Münster, 48149 Germany*
[be] *Utrecht University, Department of Physics and Astronomy, PO Box 80000, Utrecht, 3508 TA Netherlands*
[bf] *INFN, CNAF, v.le C. Berti-Pichat, 6/2, Bologna, 40127 Italy*
[1] *deceased*
[2] *also at Dipartimento di Fisica, INFN Sezione di Padova and Università di Padova, I-35131, Padova, Italy*

E-mail: rbruijn@nikhef.nl, marco.circella@ba.infn.it, vivolo@na.infn.it, km3net-pc@km3net.de

ABSTRACT: The optical module of the KM3NeT neutrino telescope is an innovative, multi-faceted large area photodetection module. It contains 31 three-inch photomultiplier tubes in a single 0.44 m diameter pressure-resistant glass sphere. The module is a sensory device also comprising calibration instruments and electronics for power, readout and data acquisition. It is capped with a breakout-box with electronics for connection to an electro-optical cable for power and long-distance communication to the onshore control station. The design of the module was qualified for the first time in the deep sea in 2013. Since then, the technology has been further improved to meet requirements of scalability, cost-effectiveness and high reliability. The module features a sub-nanosecond timing accuracy and a dynamic range allowing the measurement of a single photon up to a cascade of thousands of photons, suited for the measurement of the Cherenkov radiation induced in water by secondary particles from interactions of neutrinos with energies in the range of GeV to PeV. A distributed production model has been implemented for the delivery of more than 6000 modules in the coming few years with an average production rate of more than 100 modules per month. In this paper a review is presented of the design of the multi-PMT KM3NeT optical module with a proven effective background suppression and signal recognition and sensitivity to the incoming direction of photons.

KEYWORDS: Water-Cherenkov; neutrino detectors; neutrino telescopes; PMT

## Contents

## 1. Introduction

KM3NeT is a neutrino telescope located at the bottom of the Mediterranean Sea. With the ARCA and ORCA[1] detectors of the telescope scientists will search for neutrinos from distant astrophysical sources and study the fundamental properties of neutrinos, respectively. The extendable neutrino detectors comprise large volumes of seawater instrumented with arrays of detection units, vertical stringlike structures equipped with light sensitive modules – the optical modules, described in this paper (Figure 1). They measure the Cherenkov light induced in seawater by secondary charged particles from interactions of neutrinos of all flavours. The trajectories of the charged particles through the detectors are reconstructed by combining an accurate measurement of the time of arrival of the photons and the position of the optical modules (Figure 2).

At the seabed of the ARCA and ORCA sites, extendable networks of electro-optical deep sea cables and junction boxes connected to a local onshore control station are being built. Both the ARCA and ORCA sites are nodes in larger networks for marine science research[2]. The ARCA array of optical modules is gradually being built and connected to the junction boxes in the seafloor network at about 90 kilometres offshore Portopalo di Capo Passero, Sicily (Italy), at a depth of about 3500 m. The configuration of the array is optimised for the detection of high-energy neutrinos from cosmic sources. The targeted instrumented volume of ARCA is about one cubic kilometre with the option to grow further. The optical modules in the array are distributed in the seawater volume with an average horizontal distance of about 90 m and a vertical distance of about 36 m with the lowest modules at about 70 m above the seabed. At the end of 2021, 162 modules in 9 vertical detection units are installed and the construction of the next 54 detection units is proceeding. The goal is to install more than 4100 modules in 230 detection units [1].

[1] ARCA and ORCA: Astroparticle and Oscillation Research with Cosmics in the Abyss, respectively.

[2] European Multidisciplinary Seafloor and water column Observatory, http://emso.eu/

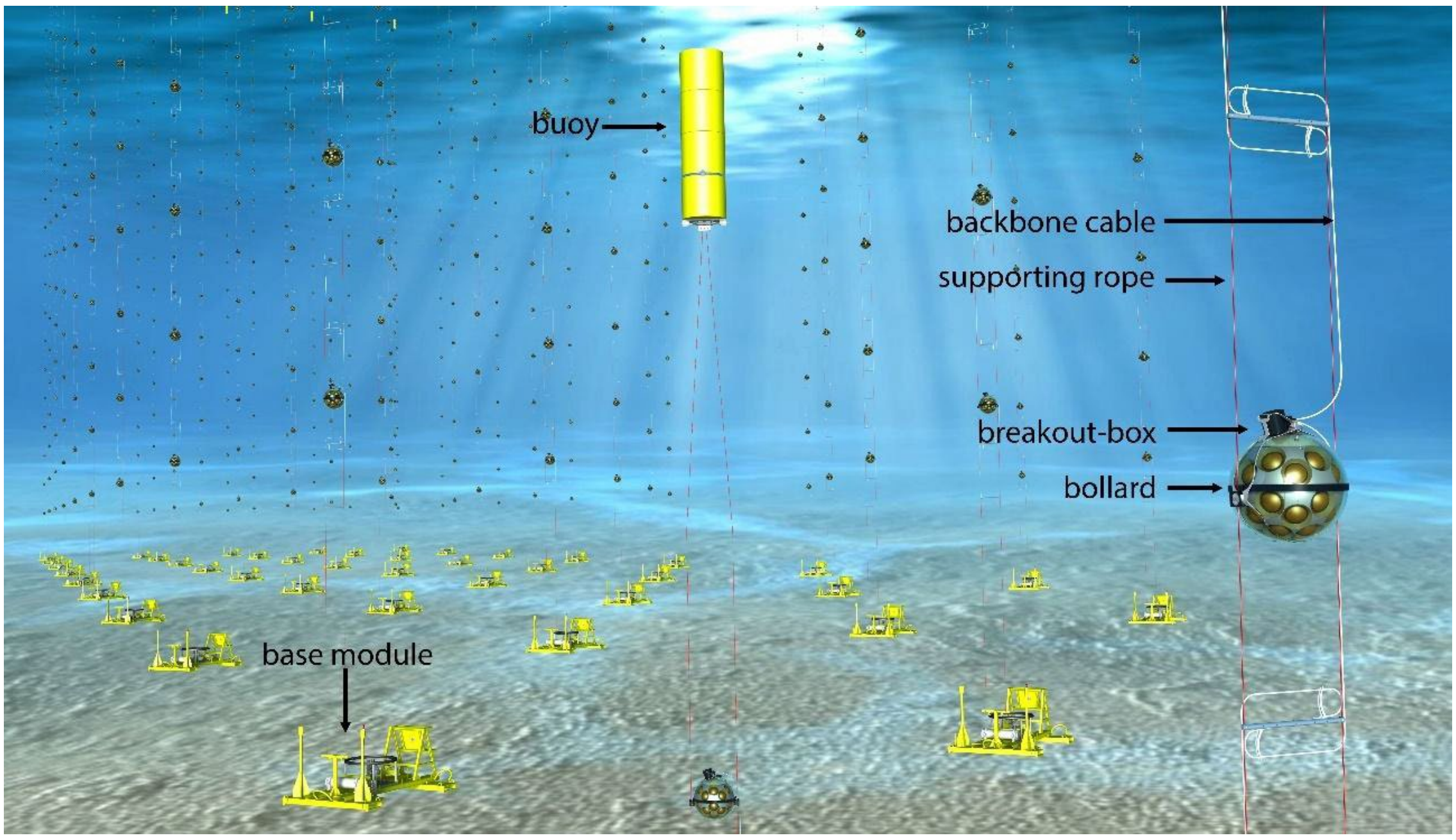

**Figure 1** Rendition of the ARCA detector showing the detection units equipped with KM3NeT optical modules, anchored to the seabed. Using bollards, the modules are attached to supporting ropes of the detection unit. Buoys at the top provide additional pull to keep the structures upright in the case of strong sea currents. An electro-optical backbone cable runs the full length of the unit with a breakout at the level of each optical module. Via a base module on the anchor and a cable running over the seabed, the detection unit is connected to a junction box in the seafloor network of the deployment site (not visible in the picture). The scale is indicative. In reality daylight does not reach the depths at which the KM3NeT neutrino telescopes are installed. The ORCA detector has a similar design.

In parallel, the array of the ORCA detector is under construction at about 40 kilometres offshore Toulon (France), at a depth of about 2500 m. The configuration of ORCA is optimised for the study of neutrino oscillations using neutrinos created in the atmosphere of the Earth by cosmic particles. Its array of optical modules will span a seawater volume of about 7 megatonnes with horizontal spacing between the modules of about 20 m and vertical spacing of about 9 m. At the end of 2021, 180 optical modules, integrated in ten units, are operational and the construction of 23 additional detection units is proceeding. For ORCA, the target is to install a total of more than 2000 optical modules in 115 detection units [1].

The optical modules applied in the ARCA and ORCA detectors have a novel design. For the first time in a neutrino telescope, the optical module design was modified from a glass sphere comprising a single large photomultiplier tube (PMT) to one with the same diameter housing not only 31 small PMTs, but also calibration devices and the full front-end and readout electronics.

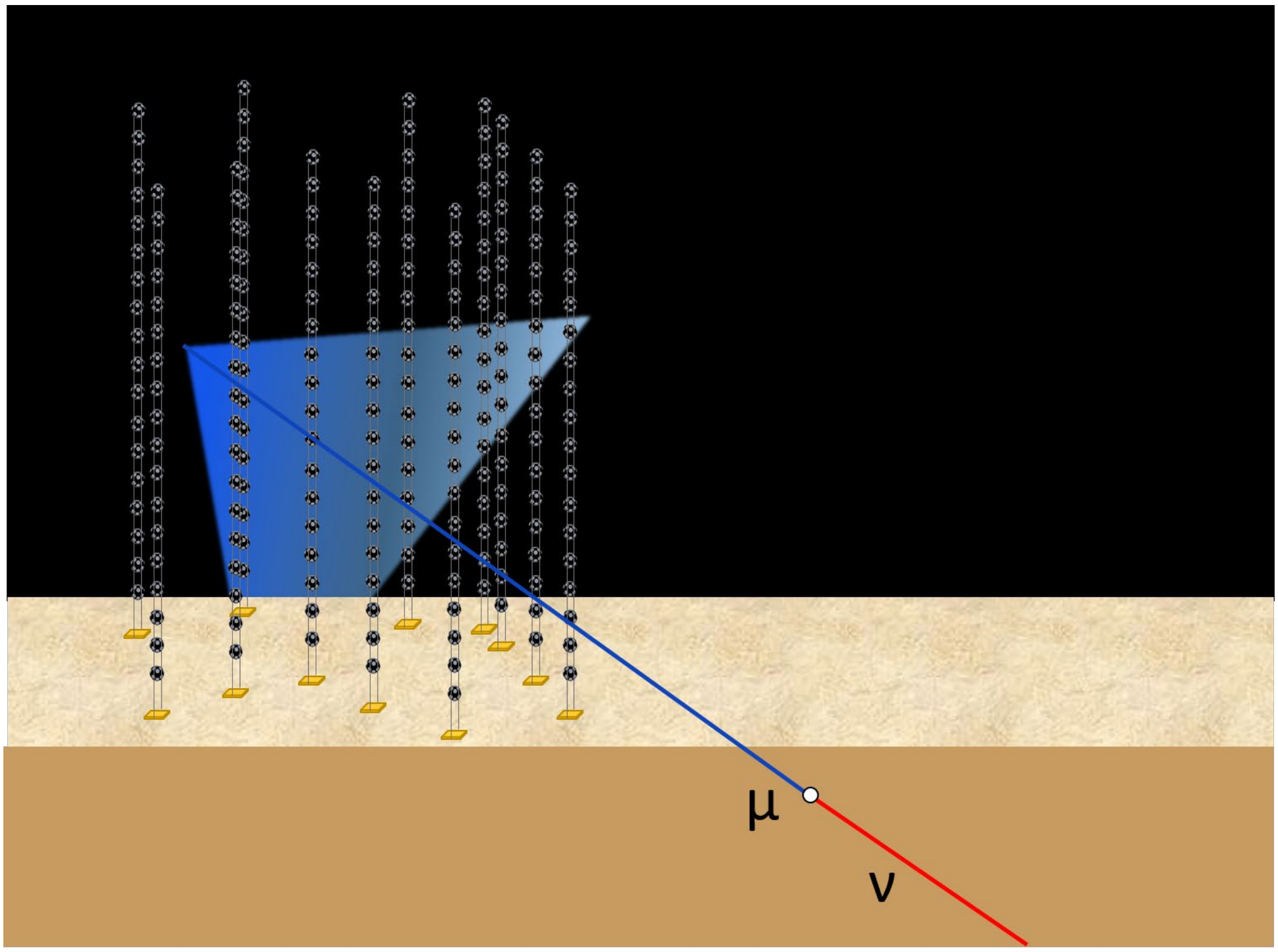

**Figure 2** Illustration of the detection principle of the KM3NeT telescope with the trajectory of a muon (blue) from a collision (dot) of a neutrino (red) with matter below the telescope. The cone of Cherenkov light emitted along the path of the muon in the instrumented volume is drawn. The direction of the neutrino is reconstructed from the time of arrival of the Cherenkov light on the optical modules and the position of the modules.

It is a design concept that can be applied in other detectors. Technical details of the major components and their qualifications have been published separately [2]-[13]. In this paper, a review is presented of the full multi-PMT design of the optical module and the distributed production model for the KM3NeT telescopes.

## 2. Design considerations

The ANTARES telescope in the Mediterranean Sea was the first to demonstrate the feasibility of operating a neutrino telescope in the deep sea [14]. The KM3NeT telescope is the next generation with a new technical design taking advantage of the experience of 15 years of ANTARES operation. At the start of the design, several requirements were formulated after an evaluation of the design of ANTARES.

Scientifically, the requirements for KM3NeT were to provide efficient detection of neutrinos with energies from GeV to PeV with a high angular resolution during an operational lifetime of at least 15 years [1],[2]. Technically, the small interaction cross-section of neutrinos and the magnitudes of relevant fluxes impose the requirement to instrument volumes of seawater of a few megatonnes for ORCA and about one gigatonne for ARCA to meet their respective primary scientific goals. The long absorption length of 70 m for photons with wavelength of 440 nm of the Mediterranean seawater [15] allows for a sparse instrumentation. The low scattering probability for optical photons, with an average scattering length in seawater of more than 100 m for blue light, can be exploited if the incoming direction of photons can be determined at optical module level and, in particular, when their arrival times on the photocathode surfaces are measured with nanosecond accuracy.

Cost-efficiency and scalability are important parameters for the realisation of a telescope of the scale of KM3NeT. Cost-efficiency was implemented by using as much as possible off-the-shelf components which companies were able to offer at competitive prices. Where necessary, when companies could not deliver at affordable prices or commercial continuity could not be guaranteed, custom-designed solutions were applied. Custom-designed solutions have been made available to companies to be applied in their portfolio. Scalability to production of a large number of optical modules is fostered by the uniformity of the modules, which allows for distributed module assembly sites with uniform quality control protocols and eases distribution of the modules over the ARCA and ORCA detectors.

At the start of the design phase, it was soon realised that the use of PMTs was still the most effective option for photodetection. Also the pressure-resistant glass spheres, which are standard items in the instrumentation for marine science research, were identified as the most reliable transparent containers. The glass of the sphere protects the sensitive equipment from the environmental conditions of high pressure in salt water, ensuring high reliability for long-term continuous operation. Traditionally, the solution for light detection in neutrino telescopes was to equip the transparent glass spheres with one large-area photomultiplier tube with diameters of 8 to 10 inches [14],[16],[17]. In the design of the KM3NeT optical module, the glass sphere is equipped with a set of 31 3”-photocathode PMTs - with approximately the same photocathode area as three 10” PMTs - of which the signals are individually processed. The concept was proposed as early as 2003 [18] and was further developed into a technically feasible and scalable implementation. The segmentation of the photocathode area provides each optical module with sensitivity for the incoming direction of the detected photons, and, in combination with the nanosecond timing accuracy, an effective tool for the reduction of background from light induced by $^{40}$K decay and bioluminescence in seawater. The segmentation allows counting the photons arriving on the module, which is not straightforward in the case of single large PMT. Also, the impact of a failing PMT on the performance of the telescope is lowered as the module can still be operated efficiently with fewer PMTs.

The design of the front-end electronics and data acquisition chain has to take into account the large counting rates per PMT, in the order of 7 kHz [3], which stem from the decay of ambient $^{40}$K and bioluminescence in the seawater. In order to allow for sophisticated algorithms to filter the signal from the background, the information concerning all detected photons is sent to the onshore control station. Encoding the PMT photon measurements with only the arrival time and pulse-width (time-over-threshold; ToT) reduces the required bandwidth still providing sufficient information. The reason is that a nanosecond time measurement accuracy allows for resolving the time structure of the photon flux with minimal distortion thanks to the long scattering lengths of the seawater.

An important mechanical requirement for the design was to minimise the number of feedthroughs in the telescope in order to minimise the risk of water leaks. By fitting the PMTs in a single glass sphere together with the calibration devices and electronics, the number of feedthroughs in the telescope could be reduced significantly compared to ANTARES.

In summary, the advantages of the multi-PMT design with advanced electronics, optical solutions for long-distance communication and cost-efficient calibration devices for the KM3NeT optical module are manifold:

- A projected photocathode area of about 1300 $cm^2$ in each sphere, which is about three times the area of a single 10" PMT, allowing a sparse distribution of optical modules in the detection volumes;
- An almost uniform and extended angular coverage of the telescope with a field of view above the horizon;
- Sensitivity to the incoming direction of detected photons;
- Good photon counting performance;
- Good position and timing calibration;
- The possibility to define local triggers (implemented onshore) based on the pattern of PMT signals
- A mechanical infrastructure of the detection unit with a small number of pressure housings and barriers as well as electronics, allowing for a significant cost reduction at parity of detector performance;
- Uniformity of the most important component of the detectors, which allows for reliable production and eases scientific analysis.

The combination of a nanosecond measurement of photon arrival times, 10 cm position accuracy ensured by an acoustic position calibration, photon flux estimation by counting the number of hit PMTs and information on the direction of incoming photons allows in the ARCA detector for a median angular resolution of 0.1 (0.055) degrees for muon tracks at 10 TeV (10 PeV) and better than 2 degrees above 100 TeV for electromagnetic cascades, while obtaining an energy resolution of about 30% for tracks and 5% for cascades at 100 TeV [19]. In case of ORCA, the achieved performance allows for a sensitivity to determine the neutrino mass ordering with 4.4 $\sigma$ if normal ordering is true, or with 2.3 $\sigma$ if inverted ordering is true with 3 years of data [20].

## 3. Technical implementation

The implementation of the design of the multi-PMT optical module resulted in a standard 0.44 m diameter pressure-resistant glass sphere with a dense packing of sensors for photodetection; devices for position and timing calibration; and the associated electronics for electrical power, readout and data acquisition, monitoring and long-range communication with the onshore control station. For illustration, a photo of an optical module after assembly is shown in Figure 3 and a rendition of the module integrated in a detection unit is shown in Figure 4. The components in the optical modules were selected to work at operational conditions of low power consumption. As a result, the electrical power consumption of the optical module could be kept at a level of 7 W.

In the following sections, details of the multi-PMT optical module design will be described with reference to the exploded view of the optical module shown in Figure 5. As an illustration, in Figure 6 a picture of the main components is shown.

### 3.1 Photodetection

*The photomultiplier tube*

For implementation in the optical module, various PMTs with comparable performance are available on the market. Because of funding arguments and adherence to national and international rules of competition, PMTs were procured in batches of several thousands. In addition, the stepwise procurement gives the possibility to profit from industrial performance improvement. For the first batches, Hamamatsu delivered the R12199-02 3 inch, which is shown together with the base in Figure 7. The model adopted for the following batches is the Hamamatsu R14374, with a slightly improved performance, in particular for the transit time spread. The PMTs have a convex bialkali photocathode, with a diameter of 80 mm, and a 10-stage dynode structure. The performance of the R12199-02 and the comparison with the requirements set by the KM3NeT Collaboration are reported elsewhere [5]. During integration of the optical module, polished metal rings (component L in Figure 5) are placed at an angle of 45 degrees around the head of PMTs providing a 92% reflectance for photons in the wavelength range 375-500 nm. The recessed placements of the rings exploits the convex shape of the sensitive area of the PMT. By doing so, the photon acceptance is increased by 20-40%, with most of the gain attained in the forward direction [6]. The PMTs are operated with a negative high voltage (HV) on the photocathode, placement allowing simpler control and signal digitisation electronics on the PMT bases. An insulating coating is applied to the outside of the photomultiplier tubes and on the PMT bases to prevent electrical discharge between the PMT and its surroundings [7]. To minimise the effect of ageing and thus maximise the lifetime of the PMTs, given the high-counting rate typical of the deep-sea environment, a relatively low nominal gain of $3 \times 10^6$ is chosen. The operational gain is obtained by tuning the HV of each PMT in the lab and *in situ* during operation of the telescope. As the information on the PMT pulse is reduced to the time of crossing a threshold together with the pulse duration – the time-over-threshold technique – the tuning is done by considering the distribution of pulse-widths of each PMT and setting the average to a value corresponding to the required gain [5]. PMTs are indicated with letter M in Figure 5.

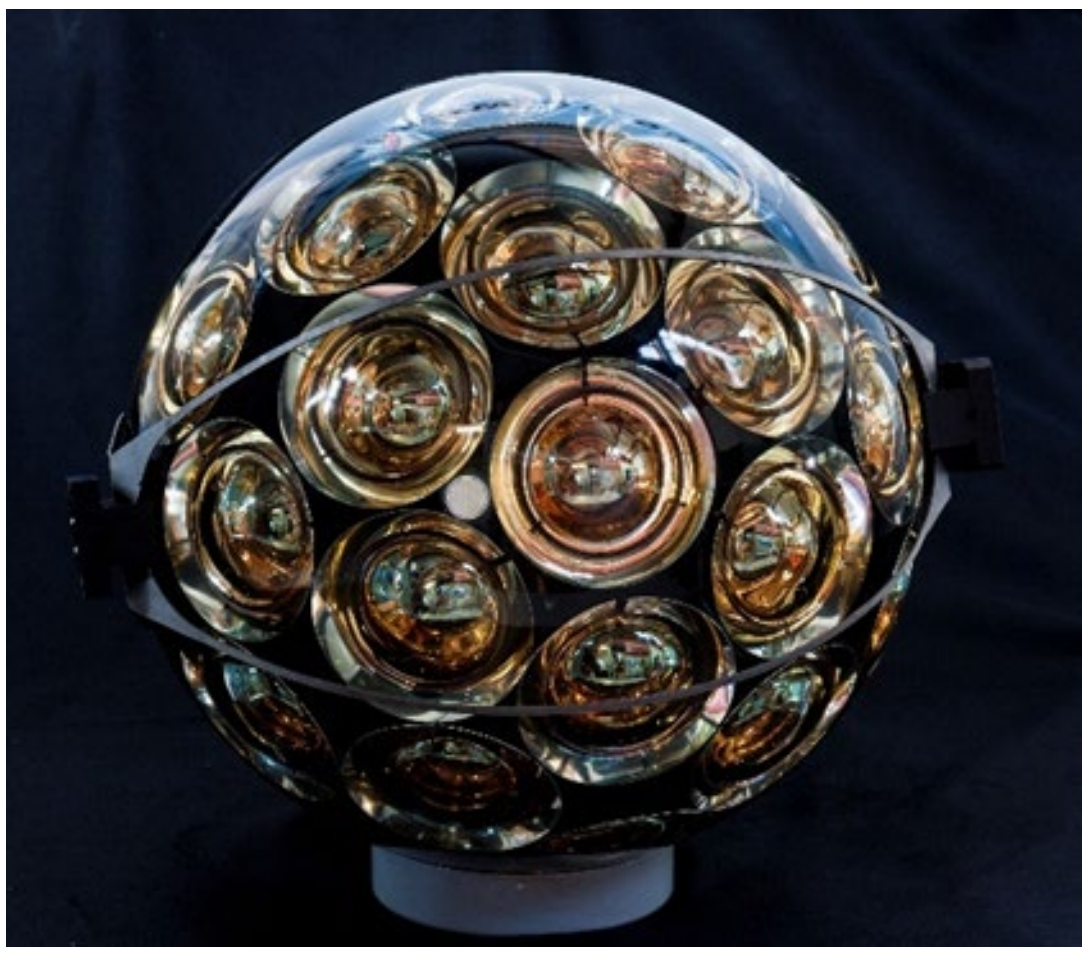

**Figure 3** Bottom view of an optical module, illustrating its large effective and segmented photon detection area. On the left of the middle PMT, the acoustic piezo sensor used for positioning measurements (see section 3.2) is visible. The titanium collar required for mounting in a detection unit is also visible.

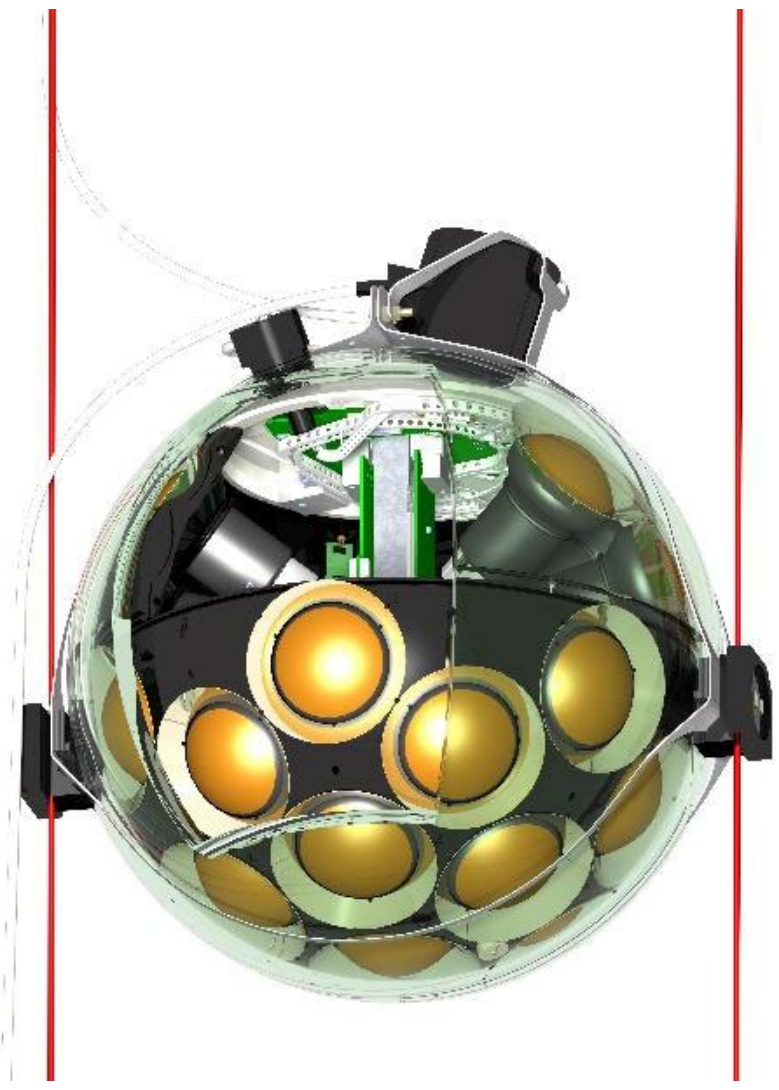

**Figure 4** Rendition of an optical module as mounted with bollards attached to the supporting ropes of a detection unit. For illustration purposes, several parts are cut out. The breakout-box which connects to the backbone cable of the detection unit (see section 3.4) is attached to the top of the module.

*The PMT base*

Each PMT has its own individual custom-designed, very low power HV base with integrated amplification and adjustable discrimination [8]. The boards have small formfactor connectors (SAMTEC 0.80 mm Tiger Eye Micro Terminal Strip) and flat kapton cables which are integrated in the printed circuit board (PCB) and carry the 3.3 V power voltage to the base together with an Inter-Integrated Circuit (I2C) communication bus and a 100 Ω LVDS (Low-Voltage-Differential-Signal) output. High voltage is generated directly on the base by a Cockroft-Walton (CW) circuit which is controlled by a custom-designed Application-Specific Integrated Circuit (ASIC) named CoCo [9]. The ASIC controls the frequency at which the power from the 3.3 V supply is pulsed inductively in the CW circuit using feedback from the HV circuit. The analog PMT pulse passes through a charge amplifier and is then digitised by another, mixed signal, custom-designed ASIC called PROMiS (PMT Read Out Mixed Signal) [10]. The digitisation is done by means of a comparator which discriminates the amplified PMT signal against a tunable threshold. A LVDS circuit drives the transmission line such that the level is kept 'high' for the time that the PMT signal remains above the threshold. That time interval, the ToT, is measured by the central logic of the optical module, while the time of the leading edge of the LVDS pulse is recorded as a proxy of the photon arrival time. An I2C block in the CoCo ASIC allows setting the HV and the comparator threshold from the onshore control station. In addition, it provides read access to a one-time programmable memory which contains a unique identifier for each PMT base. The design of the PMT base results in a very low power usage of 36 mW while in operation.

### 3.2 Position calibration devices

Since the shape of a detection unit is influenced by sea currents, the positions of the optical modules in the telescopes vary. To continuously measure their positions and orientations, an acoustic positioning system is used in combination with sensors in each optical module that provide the tilt and heading [21]. The acoustic system comprises a long baseline of acoustic emitters anchored to the seabed at known positions [22], hydrophones at the bases of the detection units and acoustic piezo-sensors in each optical module. In order to achieve the required precision of 10 cm on the position, the accuracy of the measurement of the arrival time of the acoustic pulses pulses should be better than 50 μs, which is well within the precision of the timing system. The acoustic calibration system provides updated positions of the detector modules with a typical frequency of once per minute. The acoustic piezo sensor, indicated by letter R in Figure 5, is glued to the inside of the glass of the lower hemisphere of the optical module to maximise acoustic coupling. It has a cylindrical shape, with a diameter of 18 mm and a height of 11.5 mm. The analogue acoustic signals are pre-amplified and digitised on a dedicated electronics board inside the sensor housing, at a rate of about 195 kSps with 24 bit precision. The design sensitivity of the sensors is $-160 \pm 6$ dB re 1V/μPa in the range of 10-70 kHz [23]. The acoustic data embedded in the main optical module data stream are sent to the onshore control station by the central logic of the optical module. Using a dedicated computer farm in the control station, the reconstructed positions of the optical modules are stored in the KM3NeT database [24]. The measurements with the acoustic sensors are also available for Earth and sea science studies.

A compass and accelerometer positioned on a mezzanine board, provides the orientation of the optical module, and thus of the PMTs, with respect to the telescope coordinate system. Two types of boards, with comparable performance, are in use. One board contains a triaxial accelerometer STMicroelectronics LIS3LV02 and a triaxial magnetometer Honeywell HMC5843 read by a

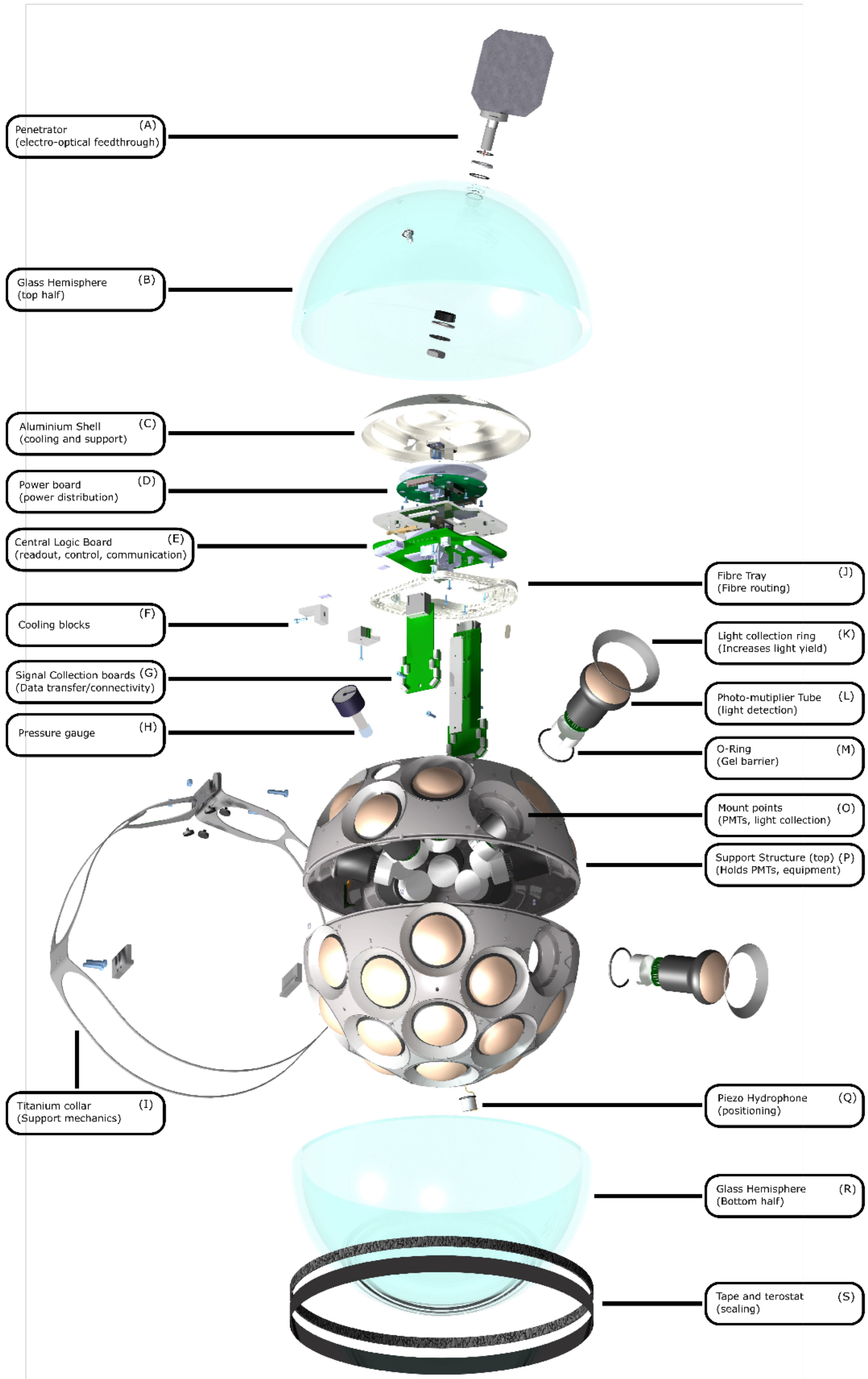

**Figure 5** Exploded view of the optical module. Lettered components are referenced in the text.

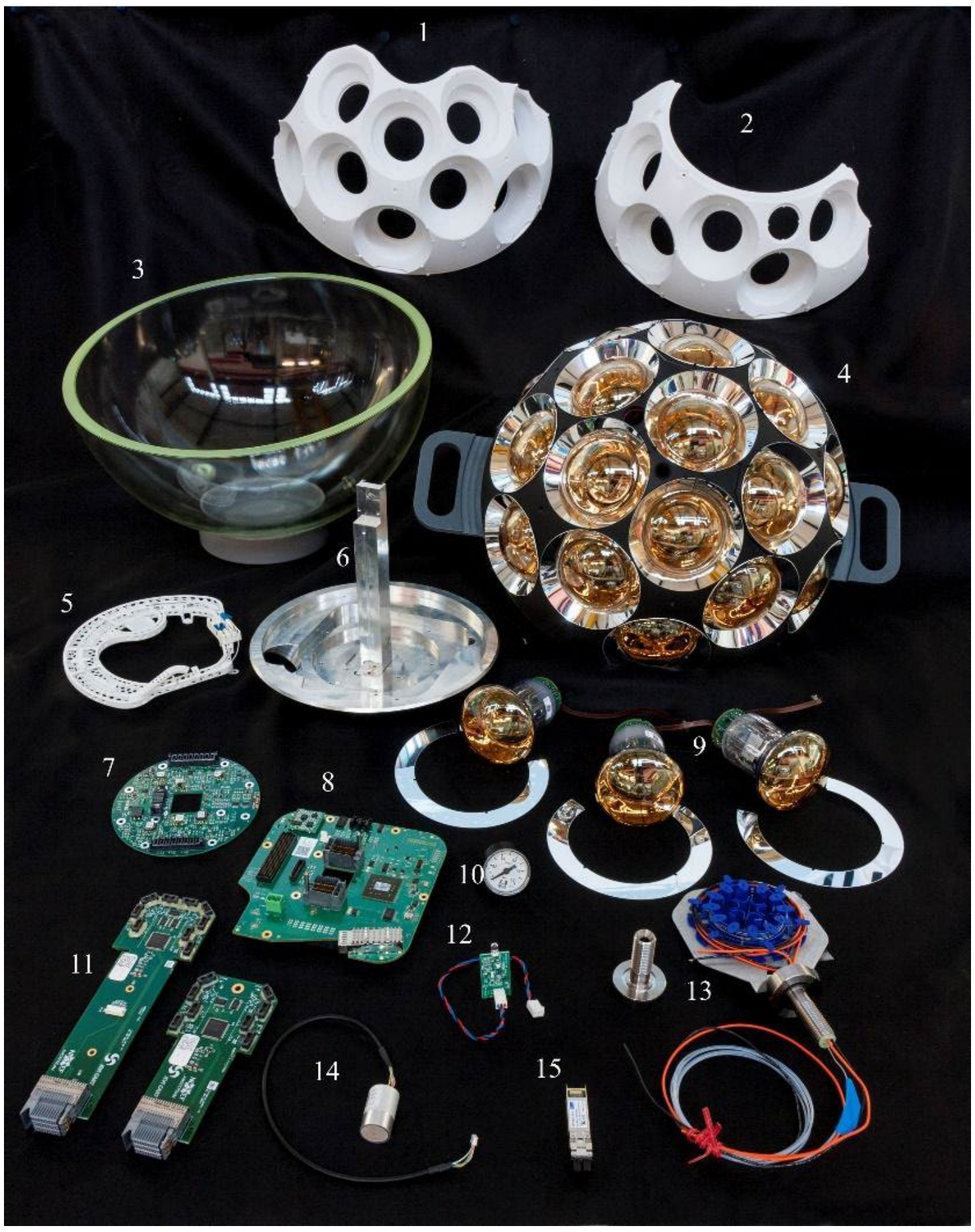

1. Section of a bottom support structure
2. Section of a top support structure
3. Glass hemisphere (bottom)
4. Bottom support structure with PMTs and light collection rings installed
5. Tray for routing of optical fibres
6. Cooling and support mechanics (shell with rod mounted)
7. Power board
8. Central Logic Board
9. (Three) PMTs with base attached and light collection rings
10. Pressure gauge
11. Signal collection boards (2)
12. Nanobeacon (led flasher) on driver board
13. Penetrator flange (left) and penetrator with temporary fibre/cable routing plate (right)
14. Piezo hydrophone
15. Laser transceiver

**Figure 6** Photo of a selection of components of an optical module.

microcontroller which communicates over an I2C bus. The other type has a single STMicroelectronics LSM303D triaxial magnetometer and accelerometer directly connected to an I2C bus. The measurements obtained by the devices are collected by the central logic board of the optical module every second and sent to the onshore control station, where they are converted to the pitch, yaw and roll of the optical module. The displacement and orientation of the optical modules can also be used to indicate changes in the sea current, providing measurements of interest for the Earth and sea science community.

### 3.3 Timing calibration

Every optical module is equipped with a 470 nm LED pulser, referred to as nanobeacon and indicated by number 12 in Figure 6, which can produce fast light pulses for timing calibration of neighbouring optical modules [25]. The intensity, the timing and the pulse frequency can be controlled remotely from the onshore control station. Notably, the times of the light pulses are well defined as their trigger is synchronised with the master clock of the telescope through the main electronics system of the optical module. The LED, together with the drive electronics, is soldered on a small electronics board, which in turn is connected to the main electronics board by means of a twisted-pair cable. The small electronics board is installed in the top hemisphere of the optical module, in optical connection to the glass of the sphere, pointing upwards. The nanobeacons are only operated during dedicated timing calibration runs. The aim of the runs is to measure the propagation time of light from the optical module in which the nanobeacon is flashing to the adjacent upper optical modules, so that the timing between different optical modules on the same or even different detection units can be calibrated. In addition, the data can be used to study the optical properties of the seawater, such as absorption and scattering length. The pulsers are typically operated at a frequency of 1 kHz, and the light intensity is adjusted such that it leads to a signal amplitude at the level of single photoelectrons in the adjacent optical module.

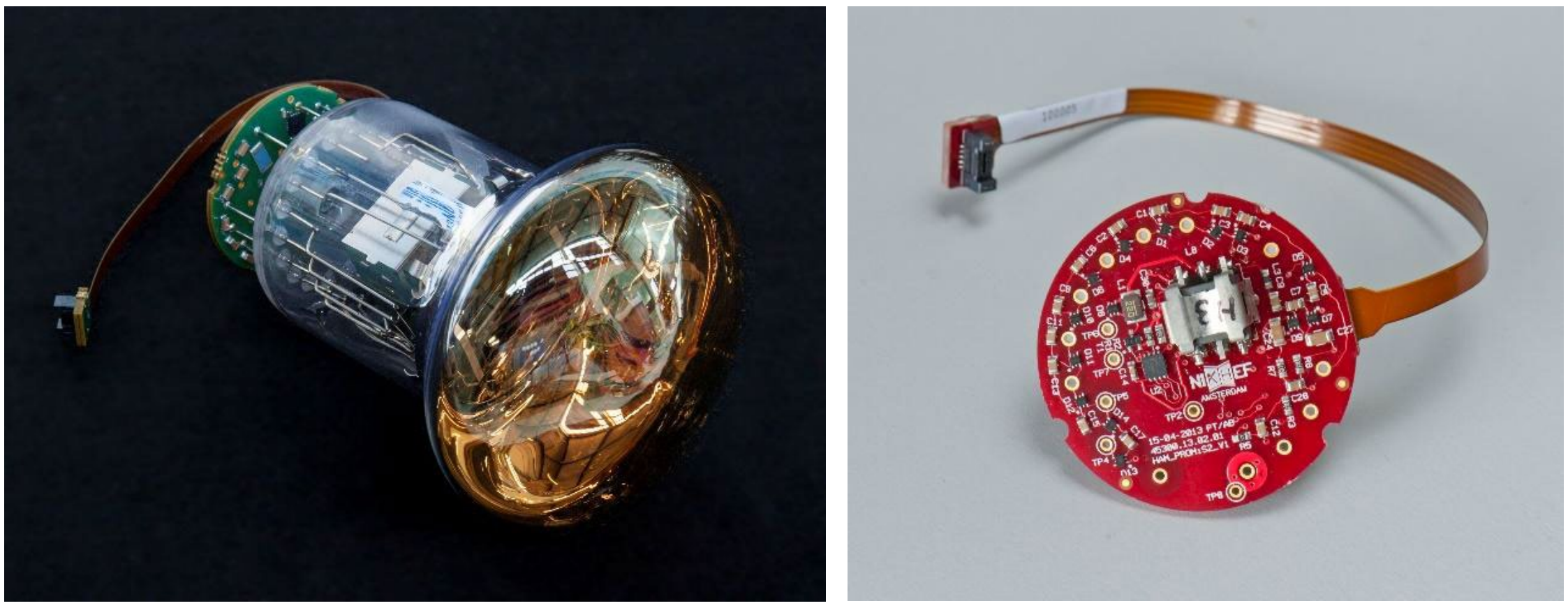

**Figure 7** (Left) A Hamamatsu R12199-02 photomultiplier tube with base attached. (Right) A photomultiplier base.

### 3.4 Data acquisition, control and time synchronisation

Central to the design of the KM3NeT telescopes and thus of the optical modules is the all-data-to-shore concept for data acquisition. In this concept, all data collected offshore are digitised and sent without reduction to the onshore control station. Using a farm of processors in the control station, data triggers and selection algorithms are applied to reduce the data volume and to extract the observables that are needed for physics analyses. The approach requires that the clocks of the optical modules are synchronised to sub-nanosecond precision. Indeed, the electronics inside the optical module serves this purpose and, in addition, controls the operation of all sensors. An in-depth discussion of the KM3NeT electronics of the readout and data acquisition system can be found in [11]-[13].

The electronics boards inside the optical module are the Central Logic Board (CLB, component E in Figure 5), two PMT-signal collection boards ('long/short octopus', components G in Figure 5), the PMT bases and the power distribution board ('power board', component D in Figure 5). External to the optical module, inside a breakout-box (featured in Figure 4) on top of the glass sphere, a DC/DC converter board converts the 375 V DC delivered via the electro-optical backbone cable of the detection unit to the 12 V DC required by the power board inside the optical module. The breakout-box provides galvanic isolation between the optical module and the rest of the detection unit, so that a failure in an optical module cannot propagate to the other optical modules in the detection unit. The DC/DC converter in the breakout-box is the only active electronic component of a detection unit that operates under ambient pressure. All components of the board have been qualified for operation at the required depths. The design of the DC/DC board is adapted to the different conditions of the seafloor networks of the ARCA and ORCA detectors.

*Central logic board*

The heart of the optical module electronics is the Central Logic Board, which controls all instrumentation inside the optical module, processes the data from PMTs and the acoustic piezo sensor and maintains the communication with the onshore control station. In addition, the CLB reads a humidity sensor, several distributed temperature sensors and the compass/tiltmeter for monitoring the operational conditions of the optical module. At power-up of the optical module, the Field Programmable Gate Array (FPGA) Xilinx Kintex 7 on the CLB is configured from an image stored on a reprogrammable Serial Peripheral Interface (SPI) memory on the CLB which contains multiple images. In the standard configuration two images are stored, of which the first one is the so-called 'golden' image to be loaded at power-up. The image performs a sequence of checks, including checks of the communication, and allows a time window for intervention before triggering the reconfiguration of the FPGA with a 'runtime' image which provides the full system functionality. Both images can be reflashed or replaced *in situ*. Safeguards have been implemented allowing for recovery from data corruption during the upload of a new image, either golden or runtime. Multiple I2C buses are connected to the FPGA for communication with the various instruments. The CLB is equipped with 32 independent time-to-digital (TDC) acquisition channels for the 31 PMTs and the acoustic piezo sensor. Those are implemented by means of de-serialisers (ISERDESE2) provided by the FPGA, while the required GHz clock is generated from a 250 MHz base clock. In this way, the arrival times of the signals from the PMT bases are time-stamped and the widths of the signals are measured, both with 1 ns resolution.

*Signal collection boards and PMT interface*

The interface between the CLB and the PMT bases is made via two 'octopus' boards, each serving the PMTs in a hemisphere of the optical module. The signal collection boards distribute the 3.3 V to the PMT bases and provide an I2C multiplexer, so that each PMT base can be addressed individually from the CLB, while routing one LVDS transmission line from each PMT to the CLB. Additionally, the long octopus board, which serves the bottom half of the optical module, is connected to the cable of the piezo sensor, carrying the required 5 V and digital communication between the device and the CLB.

*Power board*

To serve the needs of the different subsystems, the power board inside the optical module feeds fixed voltage rails of 1, 1.8, 2.5, 3.3 (2x) and 5 V and a rail that is controllable up to 30 V through a Digital-to-Analog converter (DAC). Analog-to-Digital converters (ADCs), the output signals of which are routed to the CLB, monitor the voltage and current of all rails.

*Optical communication system and time synchronisation*

A Small Form-factor Pluggable (SFP)-cage on the CLB hosts a high-power (>80 km distance) duplex single mode laser transceiver for communications over the fibre network with the onshore control station. In the current design of the detector, communication towards the optical modules is achieved via a broadcast from the control station on a common wavelength. To allow multiplexing of the data stream from the transceiver onto a smaller set of optical fibres, each optical module has an emitting wavelength chosen from the ITU[3] grid with 50 GHz spacing. The underwater network is designed for a delivery of a maximum of 72 unique wavelengths, thus allowing for four detection units with 18 optical modules each, on a single fibre. An optical add-and-drop filter is provided in the optical module to combine the two communication channels, to and from the SFP transceiver, into a single fibre. Initially, a type of transceiver with each a fixed wavelength was used. More flexibility was introduced in the construction of detection units by moving to an SFP type that has an integrated tunable Mach-Zehnder chip that allows setting the wavelength.

[3] International Telecommunication Union, https://www.itu.int/

Crucial to the KM3NeT telescope is the sub-nanosecond accuracy required for the relative timing between the measurements from different optical modules. It is implemented through the White-Rabbit protocol [26], enabling the synchronisation of nodes in an optical Ethernet network to a master clock. White-Rabbit functionality in the optical module is provided by a White-Rabbit Precision Timing Protocol Core (WRPC) on the FPGA. It consists of the gateware required for the physical White-Rabbit capable Ethernet layer together with a dedicated LM32 processor controlling the Precision-Timing-Protocol (PTP) traffic and Phase-Locked Loops (PLLs). Standard White-Rabbit relies on bidirectional point-to-point connections between nodes, while in KM3NeT a novel design has been implemented in which a common broadcast is sent from the onshore control station to the optical modules, while the uplink channels from the optical modules go to a standard, non-White-Rabbit, Ethernet switch fabric. The KM3NeT White-Rabbit PTP core is modified such that the clocks in the optical modules are syntonised (equal frequency) to the master clock, but the local time is set to the time received from the master, resulting in a fixed offset due to the propagation delays of the signals from the onshore control station to the optical modules, which will be determined by calibration.

### 3.5 Mechanical support and electronics cooling

*The glass container*

All optical module components are housed in a 0.44 m diameter Vitrovex® low-activity borosilicate glass sphere. The glass thickness is 14 mm and the spheres are rated to withstand pressures up to $6.7 \times 10^{7}$Pa. The glass of the sphere meets a light transmission requirement of more than 95% above 350 nm, thus matching the optical wavelength range of maximum sensitivity of the photomultiplier tubes. Optical contact between the PMTs and the glass is ensured by the application of the transparent two-component silicone Wacker SilGel® A/B gel. Since polymerisation of the silicone gel is sensitive to numerous materials, before being officially adopted, all components inside the optical module in contact with the gel have passed a qualification process based on compatibility tests. The glass sphere is composed of two hemispheres (components B and S in Figure 5). They are in contact at the equator per standard glass-on-glass, relying on the ambient pressure to keep the seal tight. An under-pressure of $2 \times 10^{4}$ Pa below atmospheric pressure at sea-level is applied in the sealed optical module. Such a value is a balance between constraints from transport, storage and deployment, during which the optical module must remain closed, and proper behaviour of all internal components and the gel. Around the equator of the optical module, a layer of Terostat 81 covered with Scotch wrap (components T in Figure 5) keeps water from entering at shallow depths during deployment.

*Internal support structures*

Internal mechanical structures hold the various components in the module firmly in place and provide cooling of the electronic boards. The PMTs are mounted in plastic support structures, one in the top hemisphere (component P in Figure 5) and one in the bottom hemisphere (component

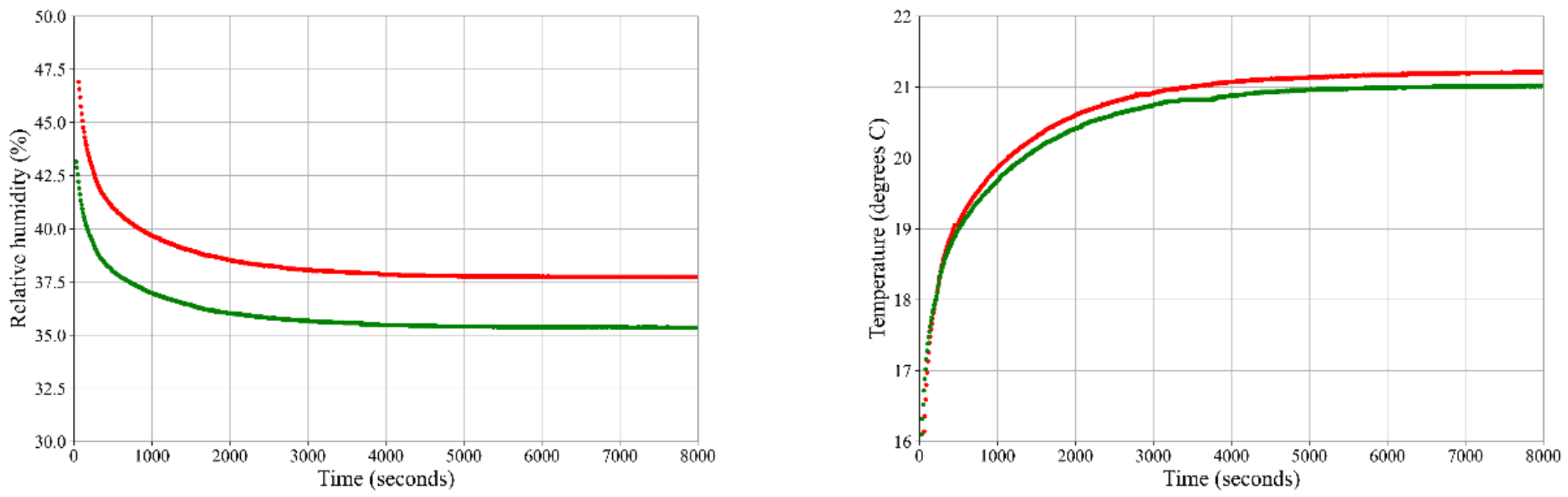

**Figure 8** *In situ* measurement of the humidity (left) and temperature (right) inside two optical modules (red and green) as a function of time since power-on. A stable condition of humidity and temperature is reached after about 2 hours.

Q in Figure 5). Until recently, the structures have been manufactured by 3D-printing using selective laser sintering of nylon or with the multi-jet fusion technology. The shape of the printed support structures includes optimisations for printing multiple parts at the same time in the limited volume of the printers. For the top structure, the 3D printing technique has been replaced by an injection moulded design using acrylonitrile butadiene styrene (ABS) for faster and significantly cheaper production. The production of the injection moulded structures includes assembling separately produced sections because a mould for the final, assembled shape is excessively more complex and expensive. The support structures are either painted black or made out of black material, depending on the manufacturing process, to minimise light reflection. The support structure defines the PMT positions and the close distance of the PMTs to the sphere glass. The PMTs are arranged in rings of 6, oriented at angles of 57.5°, 74°, 106°, 122.5° and 146.75° from the zenith of the optical module, with one PMT pointing vertically down (180°). In each ring, the PMTs are spaced by 60° in azimuth, with the PMTs in successive rings shifted by 30° in azimuth. There are 19 PMTs in the lower hemisphere of the optical module and 12 in the top one. Since sedimentation will affect primarily the top part of the module and would obscure the view of PMTs in that area, electronics and supporting mechanics are situated in the upper hemisphere of the module. The slots in the support structure for the PMTs are tapered. A silicone O-ring (component N in Figure 5) provides a tight sealing around the PMT just below its head for the 3D-printed (bottom) structures; in the moulded (top) structures, piston seals are used. The structures include some specific mounts (indicated by O in Figure 5) for the light collection rings (component L in Figure 5) around each PMT.

The top structure has mounting points with holes for the nanobeacon and a mechanical pressure gauge (component H in Figure 5); the bottom structure contains a feedthrough for the acoustic piezo sensor which is glued to the glass sphere and is sealed with an O-ring. For pouring optical gel to fill the space enclosed by the support structures and PMTs on one side and the glass sphere on the other, the structures are equipped with tubes; a dedicated system of grooves in the structure optimises the escape of air during the gel pouring.

The top part of the internal mechanical structure consists of several thermally-coupled aluminium parts which act as a passive cooling system, transporting heat away from the inside of the optical module towards the seawater. The aluminium shell (component C in Figure 5) with a primer treatment is attached to the glass by means of a thin layer of silicone gel. It houses the main electronics, provides support for the other mechanical components inside the optical module and a large surface through which heat can be dissipated to the surrounding environment. The electronics components which produce most heat, i.e. the power distribution board, the FPGA and the optical transceiver installed on the main electronics boards, are thermally coupled to the aluminium shell by means of dedicated blocks (components F in Figure 5) and heat conducting pads. The cooling system is completed by an aluminium stem which holds the cooling block for the FPGA, and attaches the signal collection boards serving the PMTs in the two hemispheres of the optical modules. The system allows for maintaining temperature conditions inside the optical modules at an optimal level for long-term reliability of the electronics. As an illustration, the temperature and humidity measured inside two optical modules operated in the sea are shown in Figure 8.

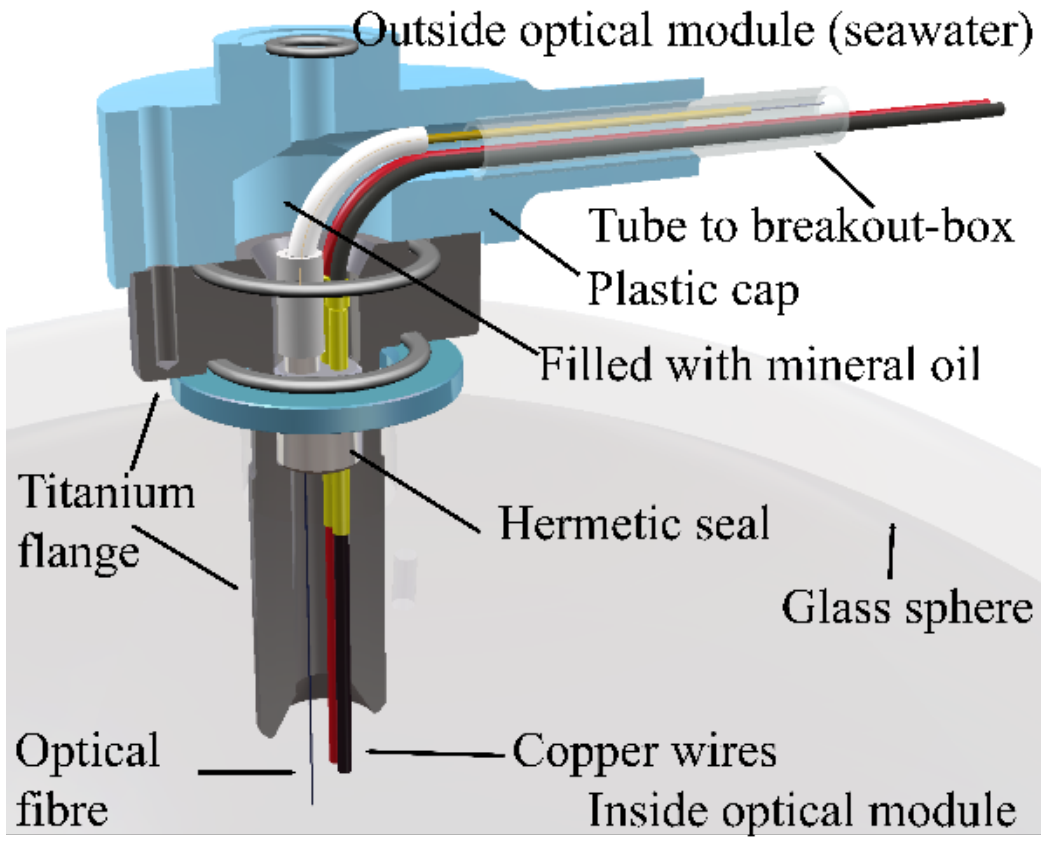

**Figure 9** Open schematic of a penetrator with cap mounted. The penetrator flange is mounted in a hole in the glass sphere, sealed with O-rings. The proprietary seal provides a hermetic throughway for two copper wires and an optical fibre.

Finally, to allow for a safe arrangement of the fibres and of the add-and-drop filter which connects the transceiver to the fibre entering the optical module, a 3D printed tray (component K in Figure 5) is fixed on the main electronic board.

*The integration in a detection unit*

In the upper glass hemisphere, holes are drilled at two locations. One hole is instrumented with a vacuum valve (component J in Figure 5) which is used when closing the optical module. The other larger hole is equipped with a penetrator (component A in Figure 5): a hermetic, pressure resistant feedthrough for copper wires and one optical fibre to connect to the electro-optical network of the telescope. The current design is developed together with Optical Fiber Packaging Ltd, using proprietary sealing technology in a titanium flange derived from an original KM3NeT design. A schematic drawing of a penetrator, including the cap that is mounted to connect it to the breakout-box via a polyethylene tube, is shown in Figure 9.

The glass sphere is enclosed in a slender titanium collar (component I in Figure 5) with polyethylene bollards for attachment at predefined positions to the synthetic Dyneema® ropes of the KM3NeT detection unit. The collar provides a mount for the breakout-box on top of the glass sphere. The shape of the collar has been designed in such a way that shadowing of the PMTs is no more than 1% in total. With all components of the PMTs and calibration devices contained inside the glass sphere, and by using minimal external support mechanics in the detection unit, the total surface area of the detection unit facing the sea current and thus the total drag on the detection unit is minimised. As a result, the background light induced by bioluminescent life interacting with the structure is reduced.

### 3.6 Qualification

As part of a staged qualification plan towards the implementation of the telescopes, several prototype optical modules have been built and operated in the deep sea. The first of these, installed on the instrumentation line of the ANTARES detector, was operated for 10 months in 2013 at 2500 m depth [3]. By the end of 2014 a prototype detection unit, equipped with three optical modules, was installed at 3500 m depth and connected to the KM3NeT network at the ARCA site [4]. That prototype detection unit was operated successfully for one year, before it was disconnected because of the need to upgrade the submarine infrastructure at the site. In September 2021, the prototype detection unit was recovered and found to be in good condition. The campaign allowed for a functionality test of all devices in operational environmental conditions as well as for an evaluation of the *in situ* performance of the optical modules. Following the qualifications, a series of extensive lab-based checks was performed. Inspired by the MIL-STD 810F standard and based on the expected life profile of the integrated optical modules and the various levels of integration, transportation, storage, deployment and operation in the sea, the following stress tests were performed:

- damp temperature cycles, up to 50°C and 93% humidity;
- temperature shocks from 50°C to 10°C;
- vibrations over three axes with 1 mm amplitude in the 10-500 Hz range, in random mode;
- mechanical shocks.

Fully integrated optical modules were operated in dedicated test facilities with dark boxes in which the PMTs were operated at nominal voltages and their response to calibration light sources was checked.

The qualification stage was completed with a pre-production of 18 optical modules, which were integrated into the first detection unit installed at the ARCA site in December 2015. The detection unit is still operational at the time of writing. The tests provided input for the optimisation of the design of various mechanical components of the optical modules.

## 4. Production

### 4.1 Production model

For the production of the more than 6000 optical modules for the KM3NeT telescope, a distributed production model has been established. The integration procedure has been standardised to achieve the baseline production rate required by the KM3NeT Collaboration with a moderate size laboratory space, a dark box (or dark room), relatively inexpensive equipment and modest personpower. In this way, assembly staff from the many institutes and groups in the KM3NeT Collaboration participate in the construction of the telescope, giving them at the same time the visibility in the Collaboration that their funding authorities welcome. At the end of 2021, there are 8 optical module production sites which together have a capacity to build optical modules at a rate of about 100 per month.

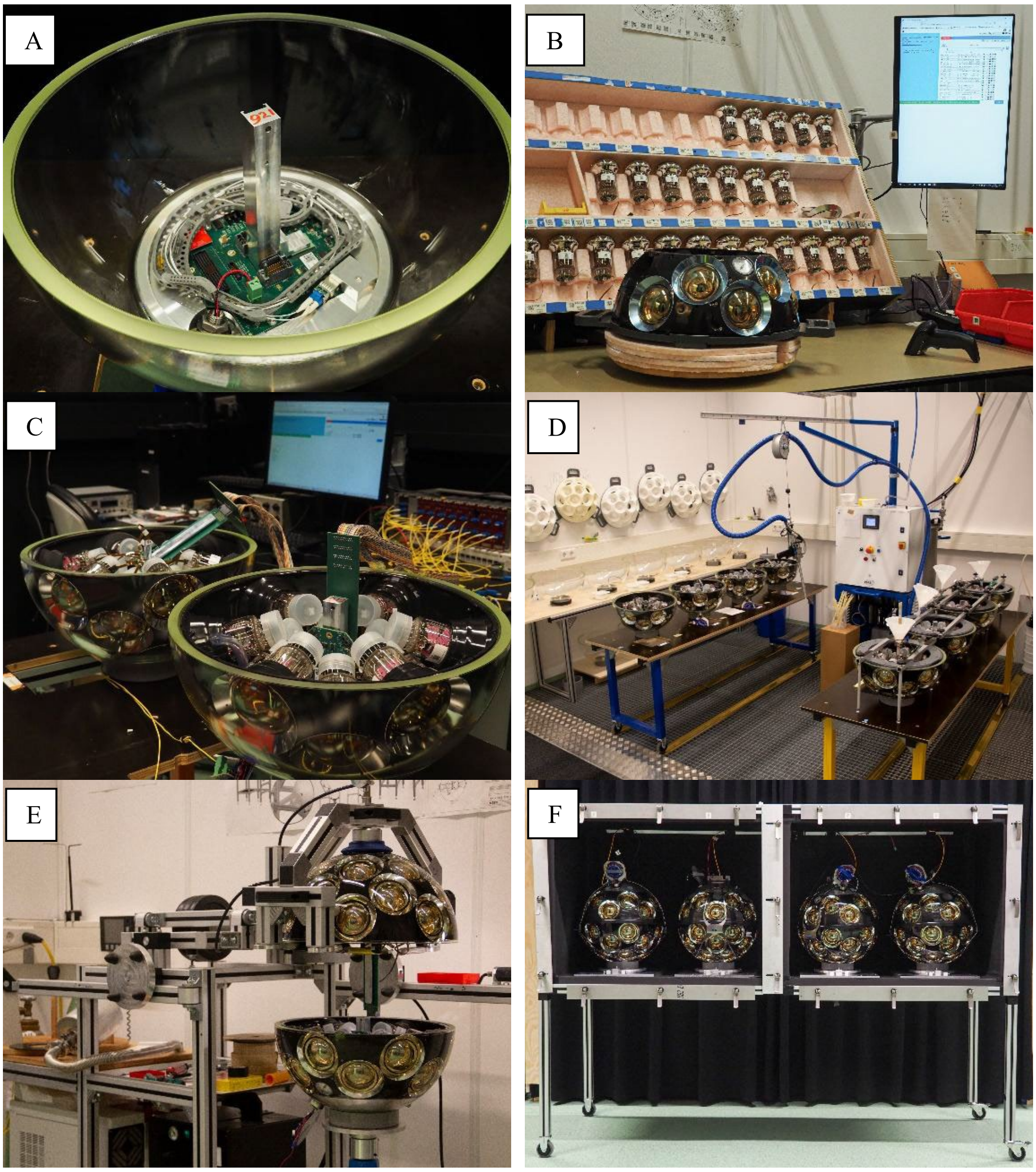

**Figure 10** Photos showing different stages of the integration of an optical module. A: Top hemisphere with cooling mechanics, electronics and penetrator mounted. B: The insertion of PMTs and light collection rings in a bottom support structure. C: Two hemispheres of an optical module connected for a functional test. D: Two top and two bottom hemispheres (visible on the right) prepared for pouring of the optical gel. The gel mixing device is also seen. E: Bottom and top hemispheres mounted in a closing station. F: Four optical modules in a moveable dark box, which includes rotation tables for compass checks and acoustic emitters.

The distributed production model comes with a cost. Procurement and shipment of the optical components to the different sites is challenging, in particular in combination with the spending requirements of the regional funds which are allocated to KM3NeT. In addition, a very high level of quality control is required to ensure that all the optical modules produced at various sites conform to the KM3NeT quality standards.

### 4.2 Integration

The integration[4] of an optical module with the dense packaging of delicate components requires a strict protocol. Each hemisphere of the modules is filled with components bottom-up, i.e. with the convex side of the glass hemisphere facing down, allowing access to the inside. This continues, as explained in the following, until the two hemispheres are completed and tested. After closing the optical module and placing the external titanium structure around the glass sphere, the optical module is ready for integration into a detection unit.

The equipment of the hemispheres starts with those components which need to be glued to the glass. In the upper hemisphere, it is the aluminium support and the cooling shell; in the lower hemisphere the acoustic piezo sensor. The stack of the electronic power board, the CLB and an electromagnetic shielding plate is mounted into the aluminium shell in the upper hemisphere and the penetrator is installed; a helium leak test assesses successful mounting of the penetrator. The copper wires from the penetrator are connected to the power board and the optical fibre is fusion-spliced to the add-and-drop filter connected to the laser transceiver. In Figure 10A a photo of the result of the actions described above is presented. A power-up test is performed to check that the electronic boards are well installed and operational. After installing the nanobeacon and pressure gauge, the PMT structures are equipped with PMTs, and light collection rings are installed (see photo in Figure 10B). The pigtail cables from the PMTs and the cable from the piezo sensor and the nanobeacon are connected to the octopus boards. A functional test is performed by loosely the two halves by means of a special extension cable, as featured in Figure 10C. The functional test is the last check and the last chance of replacing faulty components. Following this, integration is resumed by pouring optical gel to fill the space between the PMT support structures and the internal surface of the glass. The gel pouring station, with a set of two bottom and two top hemispheres prepared, can be seen in Figure 10D. After thoroughly cleaning the glass contact surfaces, the optical module is closed and sealed. The operation is performed by lowering the bottom hemisphere onto the top one using a dedicated tool that allows for rotation and translation of a hemisphere, making use of a suction cup. In Figure 10E a bottom half and top half mounted in the closing station are shown. During the operation, when the two hemispheres are at the appropriate distance, the octopus board of the bottom hemisphere, which is mounted on a sliding support, is plugged in the CLB. After installing small springs that act between the top and bottom support structures, the two hemispheres are put in contact and an internal under-pressure is induced by means of a pump connected to the valve in the top hemisphere. Finally, the contact area between the two hemispheres at the optical module equator is sealed. The optical module assembly is completed with the installation of the external titanium collar. A final acceptance test is performed by placing the optical module in a dark box or a dark room to allow for operation of the PMTs; such a test provides the input for the decision whether to accept or reject the optical module against a set of predefined criteria and allows a first calibration of the assembled optical module. A module dark box with a capacity for four optical modules is shown in Figure 10F.

### 4.3 Quality control

The integration of an optical module is an elaborate process, requiring expertise in mechanical mountings, handling and installation of electronic boards and optical components, and test procedures. Handling of electronic components is done on ESD[5]-protected workbenches and under controlled environmental conditions. In order to minimise the problems which may occur at the level of integration, all components go through checks, of which the details depend on the component, before being accepted and shipped to an integration site. Such screening includes full functionality tests and environmental stress screening of all electronic boards, calibrations of the different instruments and qualification under pressure of all penetrators. The checks are done at those institutes that have the relevant equipment to perform the tests, such as pressure tanks, climate chambers and calibration setups. Upon completion of the checks, the characteristics of the individual components (e.g. calibration constants) are recorded in a central database. The scheme for quality control of the PMTs includes sampling the batches of delivered PMTs in setups consisting of dark boxes equipped with readout electronics and pulsed lasers.

---

[4] The integration process and an integration laboratory are featured in a video found at https://www.youtube.com/watch?v=tzxHlLgAahE

A distributed production model requires a well-defined integration procedure, uniformity of instrumentation and tools over the multiple integration sites, a detailed planning for procurement and delivery of components, and the definition of dedicated quality assurance and control protocols to guarantee the same quality standards everywhere. All steps of the integration have been studied for protection against errors, while minimising the construction time. Deviations from the prescribed procedure are tracked through a non-conformity handling protocol established as part of the global quality plan of the KM3NeT Collaboration: whenever a problem occurs, a non-conformity report is opened and the corresponding component or assembly is temporarily removed from the production until the non-conformity report is resolved. The procedure can lead to three possible outcomes: the component or assembly is accepted by waiving the non-conformity; the non-conformity is corrected; or the component or assembly is discarded. Deviations from the expected results during the functional tests of the optical modules are also tracked through the non-conformity management process.

All components used in the construction of the KM3NeT detectors are registered in a central database with a unique product identifier (UPI). The UPI is also physically attached to each component, preferably by means of a QR code. The history of the components is tracked in the database and where required, e.g. for the PMTs, settings or calibration constants are stored. The values can be then retrieved from the database during detector operation and data analysis.

All integration steps are performed with the assistance of the dedicated KM3DIA integration software, of which the main functions are to indicate the right sequence of the operations, to log all relevant information and to register all optical module integration details. A hand-held scanner is used to read the QR codes of the components being integrated into an optical module so that the software can record the association. During the integration process, the information is stored in a local database for all optical modules and is transferred to the central database of the experiment once an optical module is completed so that the information is available during further integration steps and detector configuration and operation.

---

[5] Electrostatic Discharge

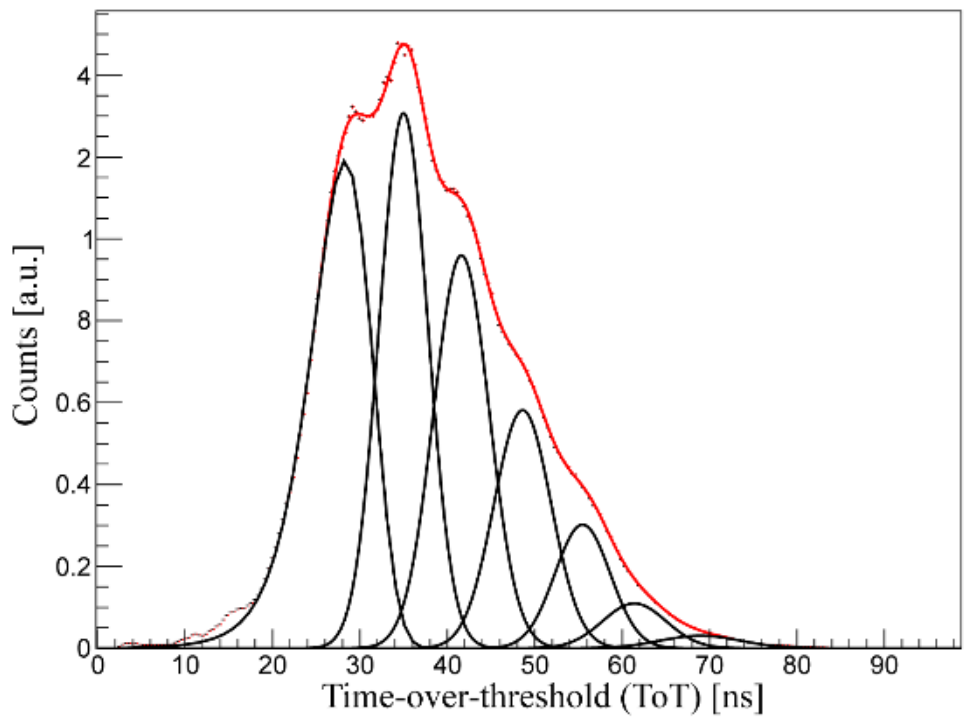

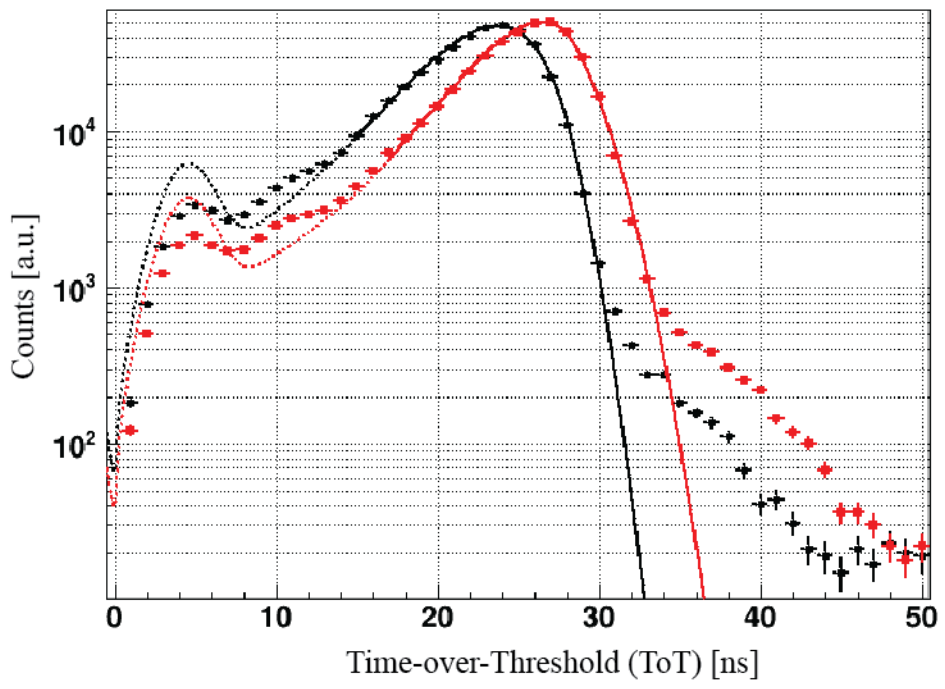

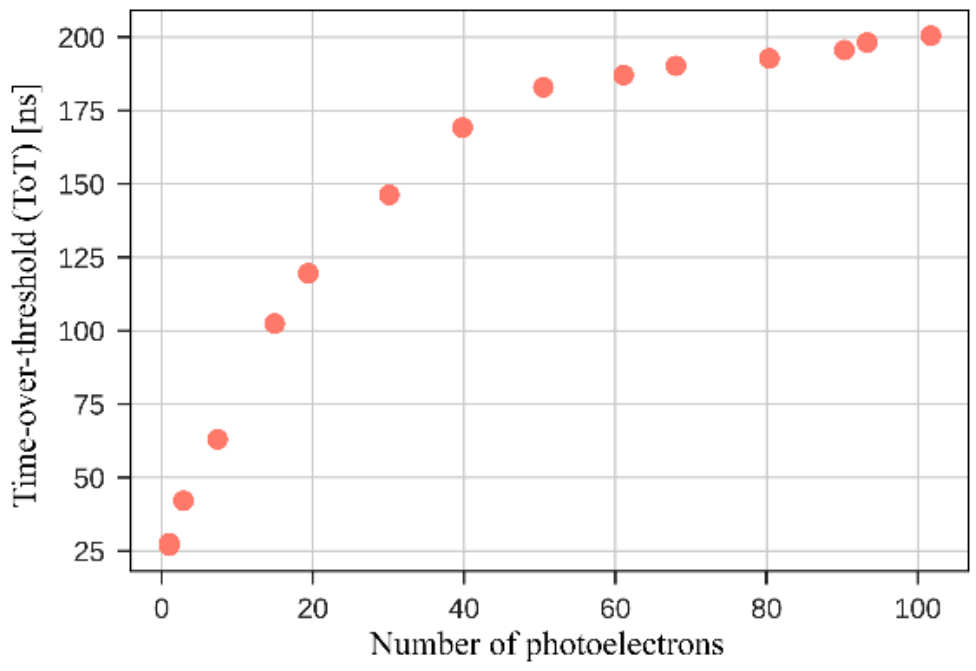

**Figure 11** : (Top left) Lab measurement of time-over-threshold response of the Hamamatsu R12199-02 PMT to different numbers of photo-electrons [29]. The black lines are model fits for increasing numbers of photo-electrons, starting at 1 photo-electron, the red line is an overall fit. (Top Right) *In situ* measured time-over-threshold distribution with single photo-electron model fits for a single PMT operated at different high voltages (black: -1130 V; red: -1180 V, solid lines indicate fit region) [30]. (Bottom left) The average time-over-threshold as a function of the number of photo-electrons [31].

## 5. Performance

The potential to enhance the physics performance of an underwater neutrino telescope, stemming from the KM3NeT multi-PMT design, has been validated early during the aforementioned qualification phase and subsequent operation of the first detection units of ARCA and ORCA.

At the end of 2021, eighteen detection units comprising in total 324 optical modules – 10.044 PMTs - are operated, including the first full ARCA detection unit deployed already in 2015. The data collected *in situ* have been already used to illustrate the performance of the multi-PMT optical modules. The rates of multifold signal coincidences on the optical modules are monitored continuously and show reproducible behaviour. With a single optical module, the signal from atmospheric muons can be distinguished from the background originating from $^{40}$K decay and bioluminescence, by exploiting multi-PMT coincidences [3],[4]. High-multiplicity coincidences allow for a measurement of the depth-dependence of the atmospheric muon flux [27], while a lower number of coincidences originating from individual $^{40}$K decays are used to calibrate the individual PMT time offsets in optical modules and the relative photon-detection efficiency [27]. Using data from the first six detection units of ORCA, the potential of a real-time supernova trigger that probes excesses of high-multiplicity coincidences has been shown [28].

The ToT response of the PMTs towards multiple photons has been studied in lab measurements and *in situ*. The shape of the distribution of the ToT values due to single photo-electron signals is described by an analytical model, which is fit to the data with the gain and gain-spread of the PMT as free parameters. With the model, the PMT gain can be monitored *in* *situ* and if needed adjusted by HV tuning [29]. Several measurements of the ToT response of the PMTs and models can be seen in Figure 11. The time offsets between different optical modules are adjusted by exploiting signal correlations between different optical modules when flashing the LED nanobeacons in the optical modules or by using signals from atmospheric muons passing through the detector. After *in situ* HV tuning, adjustments of only a few nanoseconds of the time offsets are needed when comparing the values with those from the calibrations done onshore using laser signals prior to deployment. The distributions the differences between the expected and measured photon arrival times for signals stemming from muons are narrow, demonstrating the good scattering properties of the water together with the accurate time calibration [32].

## 6. Concluding remarks

In this paper, the design of the multi-PMT optical module for KM3NeT has been described. For the first time for a neutrino telescope, a set of small diameter PMTs replaces the traditional large diameter PMT in the glass sphere of the module. The concentration of all components into a single glass sphere makes the KM3NeT module more than just an optical module: it is an instrument on its own. This was demonstrated already in 2013 with the first *in situ* operation of a prototype module which showed the background suppression capabilities using coincidences between photons arriving at the PMTs in a single optical module and the capacity of individual optical modules to distinguish the atmospheric muon signal [3]. The subsequent, deployed detection units further demonstrated the added value of the multi-PMT design.

At the end of 2021, more than 700 modules have been produced, of which more than 340 are already installed in the ARCA and ORCA detectors being constructed at the seabed of the Mediterranean Sea.

A total of more than 6000 modules will be assembled at 8 production sites in the KM3NeT Collaboration applying high standards of quality control.

In conclusion, the multi-PMT module provides a large photocathode area, a broad angular coverage, accurate photon arrival time and flux measurement and self-calibration, which together enhance the reconstruction capabilities of the KM3NeT detectors.

## Acknowledgments

The authors acknowledge the financial support of the funding agencies:
Agence Nationale de la Recherche (contract ANR-15-CE31-0020), Centre National de la Recherche Scientifique (CNRS), Commission Européenne (FEDER fund and Marie Curie Program), Institut Universitaire de France (IUF), LabEx UnivEarthS (ANR-10-LABX-0023 and ANR-18-IDEX-0001), Paris Île-de-France Region, France; Shota Rustaveli National Science Foundation of Georgia (SRNSFG, FR-18- 1268), Georgia; Deutsche Forschungsgemeinschaft (DFG), Germany; The General Secretariat of Research and Technology (GSRT), Greece; Istituto Nazionale di Fisica Nucleare (INFN), Ministero dell'Università e della Ricerca (MIUR), PRIN 2017 program (Grant NAT-NET 2017W4HA7S) Italy; Ministry of Higher Education, Scientific Research and Innovation, Morocco, and the Arab Fund for Economic and Social Development, Kuwait; Nederlandse organisatie voor Wetenschappelijk Onderzoek (NWO), the Netherlands; The National Science Centre, Poland (2015/18/E/ST2/00758); National Authority for Scientific Research (ANCS), Romania; Ministerio de Ciencia, Innovación, Investigación y Universidades (MCIU): Programa Estatal de Generación de Conocimiento (refs. PGC2018-096663- B-C41,-A C42,-B-C43,-B-C44) (MCIU/FEDER), Generalitat Valenciana: Prometeo

(PROMETEO/2020/019), Grisolía (refs. GRISOLIA/2018/119,/2021/192) and GenT (refs. CIDEGENT/2018/034, /2019/043,/2020/049,/2021/023) programs, Junta de Andalucía (ref. A-FQM-053-UGR18), La Caixa Foundation (ref. LCF/BQ/IN17/11620019), EU: MSC program (ref. 101025085), Spain.